\documentclass[]{raa01}            
\usepackage{graphicx,times}
\usepackage{natbib}
\usepackage{lscape}
\bibpunct{(}{)}{;}{a}{}{,}

\providecommand{\bm}[1]{\mbox{\boldmath$#1$\unboldmath}}

\begin{document}

   \title{A compiled catalog of rotation measures of radio point sources
}

 \volnopage{{\bf 2014} Vol.\ {\bf 14} No. {\bf 8}, 942--958~~
 {\small doi: 10.1088/1674--4527/14/8/005}}
   \setcounter{page}{942}

   \author{Jun Xu
      \inst{1,2}
   \and Jin-Lin Han
      \inst{1
      }
   }

\institute{
National Astronomical Observatories, Chinese Academy of
  Sciences, Beijing 100012, China; {\it hjl@nao.cas.cn}\\
\and
University of Chinese Academy of Sciences, Beijing 100049, China \\
 \vs \no {\small Received 2014 January 5; accepted
2014 May 1}}

\abstract{  We compiled a catalog of Faraday rotation measures
  (RMs) for 4553 extragalactic radio point sources published in
  literature. These RMs were derived from multi-frequency polarization
  observations.
The RM data are compared to those in the NRAO VLA Sky Survey
(NVSS) RM catalog. We reveal a systematic uncertainty of about
$10.0 \pm 1.5$\,rad~m$^{-2}$ in the NVSS RM catalog.
The Galactic foreground RM is calculated through a weighted
averaging method by using the compiled RM catalog 
  together
with the NVSS RM catalog, with careful consideration of
uncertainties in the RM data.
The data from the catalog and the interface for the Galactic
foreground RM calculations are publicly available on {the
}webpage: {\it http://zmtt.bao.ac.cn/RM/}.
 }

\authorrunning{J. Xu \& J. L. Han }    
\titlerunning{Catalog of Rotation Measures}  
  \maketitle

\section{Introduction}           
\label{sect:intro}

When a polarized signal propagates through a magnetized medium,
the plane of polarization is rotated, and this rotation depends on
frequency. This is the Faraday effect discovered by M. Faraday in
1844.  The polarization angle $\psi$ is thus equal to
\begin{equation}
\psi=\psi_0+ {\rm RM}\cdot\lambda^2\,, \label{angle}
\end{equation}
{w}here $\psi_0$ is the intrinsic polarization angle, and the
rotation of the polarization angle $\Delta\psi=\psi-\psi_0$ is
proportional to the wavelength squared, $\lambda^2$, with a rate
RM in units of rad~m$^{-2}$. This is the rotation measure (RM),
which is an integrated quantity of the product of the free
electron density $n_{\rm e}$ and magnetic field strength ${\vec
B}$ along the line of sight from the source to us, and is
expressed by
\begin{equation}
{\rm RM} =0.81\int_{\rm source}^{\rm us} n_{\rm e} {\vec B} \cdot
d{\vec l}\, . \label{rm0}
\end{equation}
The electron density $n_{\rm e}$ is in cm$^{-3}$, the magnetic
field is a vector ${\vec B}$ in units of $\mu$G, and $d{\vec l}$
is the unit vector {along} the light{ }path towards us in units of
pc. Only the component of the magnetic field along the line of
sight determines the amount of Faraday rotation.

If there is no $n\pi$ ambiguity for the polarization angles, the
RM value of a polarized radio point source can be determined by
polarization observations at two frequencies, through
\begin{equation}
{\rm RM}=(\psi_1-\psi_2)/(\lambda_1^2-\lambda_2^2)\, .
\label{rm12}
\end{equation}
Here $\psi_1$ and $\psi_2$ are the polarization angles at the
wavelengths $\lambda_1$ and $\lambda_2$.  Because of the $n\pi$
ambiguity {in} $\psi$ values, in practice polarization angles at at
least three frequencies are needed to determine RM. When
multi-frequency polarization observations are available {from} a
radio source, the slope of a linear fit to polarization angles
against wavelength squared is the RM, if the polarization angles
have been properly unwrapped to correct for the $n\pi$ ambiguity.

The RMs of many radio sources have been determined well using
multi-frequency polarization observations. In the early days,
polarization observations were carried out for strong radio
sources with single-dish radio telescopes, and measurements of
polarization angles at several frequencies were used to estimate
RMs \citep[e.g. sources in][]{skb81,bmv88}. Later, synthesis radio
telescopes {were} used for polarization observations with
excellent resolutions, so that RMs of different emission
components of radio sources could be measured separately
\citep[e.g.][]{ms96}. Recently, wideband observations have made it
possible to determine RMs from a set of measured $\psi$ values of
many channels in a single frequency band \citep[e.g.][]{btj03}, or
directly from the Stokes $Q$ and $U$ values of the channel maps
using the technique of RM synthesis \citep{bd05}.

Observations for RMs have great scientific merit{s}. RMs
of radio sources in small regions on the sky have been used to probe
the magnetic fields in galaxy clusters
\citep{hoe89,ckb01,je04,gdm+10,bfm+10}, in nearby galaxies
\citep{hbb98,ghs+05,mgs+08,mmg+12b,ghba13}, in stellar bubbles
\citep{ssf13} or HII regions \citep{hmg11,rgt12} and even supernova
remnants \citep[SNRs:][]{kim88,sim92,srw+11}, and high velocity
clouds \citep{mmg+10} in our Milky Way. RMs of radio sources behind
the Galactic disk revealed the magnetic structure in the disk
\citep{sk80,sf83,bhg+07,vbs+11}. The RM distribution over the whole
sky has been used for delineating magnetic fields in the Galactic
halo \citep{hmbb97,hmq99} and for deriving the Galactic foreground
RM \citep{ojr+12}.

Early RM catalogs of galaxies or quasars were compiled by, for
example,
 \citet{er79}{ and} \citet{ti80}. The most often
used are the RM catalogs for 555 objects by \citet{skb81} and
674 objects by \citet{bmv88}. Over the last ten years, RMs of a
large number of extragalactic radio sources (EGRs) have been derived
in many surveys, for example, the Canadian Galactic Plane Survey
\citep{btj03} {and }the Southern Galactic Plane Survey
\citep{bhg+07}. Observations of specific regions also increased the
total number of RM data, such as those for the Large Magellanic
Cloud \citep{mmg+12b}, {the }Small Magellanic Cloud
\citep{mgs+08} and the Galactic poles \citep{mgh+10}. These RMs
{are }in general well{ }determined, because the
polarization angles of many frequency channels have been used to
derive the RM values.

We have extensively searched literature {published in the} last two
decades for RM data. In Section~2, we publish our compilation of RM
data for 4553 point sources, which should be valuable for many
research projects, as mentioned above. Archival surveys and
{databases} are checked for
 possible associations of radio
sources with known objects, sometimes even with known redshifts.
\citet{tss09} have reprocessed the 2-channel polarization data
from the NRAO VLA Sky Survey \citep[NVSS,][]{ccg+98}, and obtained
RMs for {37\,543} sources. In Section~3 we will also compare the
RM values that we compiled with those in the NVSS RM catalog of
\citet{tss09}. We will show the distribution of RM uncertainties,
and derive the Galactic foreground RMs in Section~4 by using the
weighted averages of RM data. Discounting such a foreground RM is
important, for example, to calculate the intrinsic RMs of radio
sources \citep[e.g.][]{lea87,akm+98,bbh+07,sch10} and to
understand the magnetic fields in galaxy clusters
\citep[e.g.][]{ckb01,bfm+10,bvb+13}.


\begin{landscape}
\begin{table*}
\centering

\vspace{-5mm}
\begin{minipage}{120mm}

\caption{A Compiled Catalog of Faraday Rotation Measures for Point
Radio Sources  \label{tab1}}\end{minipage}

\scriptsize\setlength{\tabcolsep}{1.3mm}
\begin{tabular} {lllllrrrlllcllllll}
\hline\noalign{\smallskip}
No. &   RA(J2000)  & Dec(J2000)   &Note & GLong.  & GLat.   &  RM
 & $\sigma_{\rm RM}$  & Gd & Ref1.& Telescope&  Freq1 - Freq2
  & Nfre & Reso & OBJe & z.Obje & Ref2. & Remark\\
    &  hh mm ss.ss & 
    dd mm ss.ss &     & (\dg) 
     & (\dg) 
      & \multicolumn{2}{r}{rad m$^{-2}$} & G+ &      &          &      (GHz)     &      &  ($''$) &      &        &       &       \\
(1) &     (2)      &   (3)        &(4)  &  (5)    &  (6)    & (7)  & (8)&  (9) &  (10)   &      (11)      & (12) &  (13)& (14) & (15)   & (16)  & (17) &(18) \\
\hline
00001   &00 00 00.1  &+67 08 00   &O&   117.953&   4.760&  --411&  29&  C&  btj03  & DRAO/ST   & 1.402 -- 1.437&    4&   60&    &      &      &                    \\
00002   &00 00 31.6  &+66 52 43   &O&   117.953&   4.500&   --88&  16&  C&  btj03  & DRAO/ST   & 1.402 -- 1.437&    4&   60&    &      &      &                    \\
00003   &00 00 38.4  &+43 57 48   &O&   113.36 & --17.97~~~&   --70&   8&  B&  mmg+12a& VLA/D.Dnc & 1.365 -- 1.486&   14&   60&    &      &      &                    \\
00004   &00 01 00.91 &--25 04 51.90 &O&  ~~40.359& --78.502&    16&   5&  B&  mgh+10 & ATCA      & 1.384 -- 2.368&   64&   30&    &      &      &                    \\
00005   &00 01 10.60 &--33 29 28.90 &O&   359.465& --77.438&     8&   3&  A&  mgh+10 & ATCA      & 1.384 -- 2.368&   64&   30&    &      &      &                    \\
00006   &00 01 28.8  &+41 04 24   &O&   112.89 & --20.82~~~&   --74&  14&  C&  mmg+12a& VLA/D.Dnc & 1.365 -- 1.486&   14&   60&    &      &      &                    \\
00007   &00 01 53.37 &--30 25 08.50 &O&   ~~13.143& --78.662&     4&   2&  A&  mgh+10 & ATCA      & 1.384 -- 2.368&   64&   30& GAL& 1.3025& NED  &                    \\
00008   &00 01 55.2  &+36 22 48  &O&   111.92 & --25.44~~~&  --104&  13&  C&  mmg+12a& VLA/D.Dnc & 1.365 -- 1.486&   14&   60&    &      &      &                    \\
00009   &00 01 55.63 &--21 49 59.80 &O&  ~~55.438& --77.562&     4&   3&  A&  mgh+10 & ATCA      & 1.384 -- 2.368&   64&   30&    &      &      &                    \\
00010   &00 02 11.96 &--21 53 09.20 &O&  ~~55.357& --77.642&     6&   1&  A&  mgh+10 & ATCA      & 1.384 -- 2.368&   64&   30& GAL&      &      &                    \\
00011   &00 02 31.33 &--34 26 14.1  &O&   355.082& --77.218&    --6&   2&  A&  bbh+07 & ATCA      & 1.384 -- 2.368&    5& 15.8& RAG&      &      &  J000231--342614N*  \\
00012   &00 02 31.33 &--34 26 14.1  &O&   355.082& --77.218&     3&   6&  B&  bbh+07 & ATCA      & 1.384 -- 2.368&    5& 15.8& RAG&      &      &  J000231--342614S*  \\
00013   &00 02 45.32 &--30 28 37.50 &O&  ~~12.622& --78.829&    31&   2&  A&  mgh+10 & ATCA      & 1.384 -- 2.368&   64&   30&    &      &      &                    \\
00014   &00 02 55.61 &--26 54 47.00 &O&  ~~31.306& --79.198&     8&   4&  B&  mgh+10 & ATCA      & 1.384 -- 2.368&   64&   30& GAL& 0.0666& NED  &                    \\
00015   &00 03 01.37 &--31 18 10.10 &O& ~~~~8.483& --78.650&    11&   6&  B&  mgh+10 & ATCA      & 1.384 -- 2.368&   64&   30&    &      &      &                    \\
00016   &00 03 04.21 &--33 12 05.30 &O&   359.832& --77.922&     2&   3&  A&  mgh+10 & ATCA      & 1.384 -- 2.368&   64&   30&    &      &      &                    \\
00017   &00 03 19.2  &+43 34 36   &O&   113.77 & --18.44~~~&  --51&   8&  B&  mmg+12a& VLA/D.Dnc & 1.365 -- 1.486&   14&   60&    &      &      &                    \\
00018   &00 03 25.58 &--27 26 35.80 &O&   ~~28.487& --79.332&    24&   5&  B&  mgh+10 & ATCA      & 1.384 -- 2.368&   64&   30&    &      &      &                    \\
00019   &00 03 34.8  &+62 47 43   &O&   117.493&   0.430&  --355&  14&  C&  btj03  & DRAO/ST   & 1.402 -- 1.437&    4&   60&    &      &      &                    \\
00020   &00 03 41.05 &--31 04 00.30 &O& ~~~~9.408& --78.857&    76&  30&  C&  mgh+10 & ATCA      & 1.384 -- 2.368&   64&   30&    &      &      &                    \\
\multicolumn{18}{l}{......} \\
00031   &00 04 20.42 &--28 40 11.50 &O& ~~21.772& --79.485&    49&   5&  B&  mgh+10 & ATCA      & 1.384 -- 2.368&   64&   30&    &      &      &                    \\
00032   &00 04 22.38 &--23 07 33.20 &O& ~~50.947& --78.627&    16&  24&  C&  mgh+10 & ATCA      & 1.384 -- 2.368&   64&   30&    &      &      &                    \\
00033   &00 04 24.6  &+67 30 22   &O&   118.443&   5.045&  --134&  40&  D&  btj03  & DRAO/ST   & 1.402 -- 1.437&    4&   60&    &      &      &                    \\
00034   &00 04 28.35 &--30 57 34.40 &O& ~~~~9.674& --79.051&    16&   6&  B&  mgh+10 & ATCA      & 1.384 -- 2.368&   64&   30&    &      &      &                    \\
00035   &00 04 28.8  &+56 14 12   &O&   116.40 &  --6.04~~~&   --56&   7&  B&  mmg+12a& VLA/D.Dnc & 1.365 -- 1.486&   14&   60&    &      &      &                    \\
00036   &00 04 50.2  &+12 48 40   &O&   105.6  & --48.5~~~~~&   --17&   2&  A&  skb81  & various   &     ? -- ?    &    ?&    ?& GAL&      &      &  0002+12  P        \\
00036$\_1$&00 04 50.2  &+12 48 40   &O&   105.6  & --48.5~~~~~&   --17&  
2&  A&  bmv88  & various   &     ? -- ?    &    ?&    ?& GAL&      & bmv88&  0002+1232 P       \\
00037   &00 04 54.9  &+62 29 18   &O&   117.588&   0.100&  --323&  33&  D&  btj03  & DRAO/ST   & 1.402 -- 1.437&    4&   60&    &      &      &                    \\
00038   &00 04 55.2  &+44 27 48   &O&   114.25 & --17.63~~~&   --59&  12&  C&  mmg+12a& VLA/D.Dnc & 1.365 -- 1.486&   14&   60&    &      &      &                    \\
00039   &00 05 07.2  &+40 57 60   &O&   113.58 & --21.08~~~&   --72&  23&  C&  mmg+12a& VLA/D.Dnc & 1.365 -- 1.486&   14&   60&    &      &      &                    \\
00040   &00 05 16.8  &+55 17 12   &O&   116.34 &  --6.99~~~&   --53&  14&  C&  mmg+12a& VLA/D.Dnc & 1.365 -- 1.486&   14&   60&    &      &      &                    \\
\multicolumn{18}{l}{......} \\
\noalign{\smallskip}\hline\noalign{\smallskip}
\end{tabular}
\parbox{195mm}{
Notes: Col.~(1): Source number. Additional RM values in literature
are indicated with ``\_1'' and ``\_2'' or ``\_n'' for reference;
Col.~(2) and (3): Right ascension and declination (J2000) of a
source; Col.~(4): Note on source position: `O' stands for the {\it
  original} position from literature of RM measurements, `M' for the
{\it measured} position from the published radio maps in the RM
paper, `V' and `F' for source position taken from the NVSS or
FIRST survey databases, and `N' for source position taken from
NED; Col.~(5) and (6): The Galactic longitude and latitude of a
source; Col.~(7) and (8): RM and uncertainty; Col.~(9): Grade of
measurements; Col.~(10): Reference for RM observations; Col.~(11):
Telescope; Col.~(12): Frequency range for RM observations;
Col.~(13): Number of frequencies or channels for observations;
Col.~(14): Angular resolution for RM determination; Col.~(15):
Object type; Col.~(16): Redshift of the object; Col.~(17):
Reference of object redshift; Col.~(18): Remarks or names in
original references.  This table is available in its entirety on
the webpage {\it http://zmtt.bao.ac.cn/RM/}. A portion is shown
here for guidance regarding its form and content.}
\end{table*}
\end{landscape}

To make the RM catalog available to the wider community, we have
developed a web interface\footnote{\it http://zmtt.bao.ac.cn/RM/}
that allows users to tabulate the RM data in a selected area and
calculate the Galactic foreground RM.

\section{Compiling the rotation measure catalog}
\label{sect:2}

Faraday rotation is {an} effect{ that arises from propagation} {through} the intervening
magneto-ionized medium between the radiation source and us, as we
discussed above, and ideally can be measured through multi-frequency
polarization observations. However, the properties of polarized radio
sources\footnote{Here we define a {\it radio source} as a more or less
  independent radio emission component, while an {\it object} such as
  a quasar can produce a few radio sources, e.g. two unresolved lobes as two
  sources in addition to a compact core.} and observational
characteristics can make things complicated. For example, when a
radio source has two or three components with different RMs
\citep{lgb+11}, observations with a low angular resolution that is
not enough to resolve these components would produce a non-linear
dependency between polarization angle and $\lambda^2$
\citep{xh12,obr+12}, and hence observations at different
frequencies with different resolutions would yield different RMs
\citep{bml12}. It is also possible that a source has different
components (e.g. diffuse or compact), each having different
spectral and/or polarization properties.  Sources unresolved in
low angular resolution observations often show extended and/or
compact components with different RMs in high angular resolution
observations \citep[e.g.][]{rcg01}. Polarization observations at
different frequencies probe different depths of a source
\citep{gr84}. Therefore, properties of a source 
 (number
of components and the difference {in} their intrinsic
polarization) as well as observational parameters (resolution,
observational bands and the bandwidth) are important factors for
determinations{ of RM}.

Most extragalactic radio sources are compact cores in galaxy
centers, 
 jets or lobes from active galactic nuclei (AGNs).
Observations with high angular resolutions and at high frequencies
can always resolve jet regions{ because they} can probe deeply
into the emission cores; and the diffuse emission detected in
lower resolution observations is often resolved and cannot be
detected.
There may be a
large contribution to RM of the compact core from the medium
between{ the core of the} source and {its} edge, in addition to
RMs from intergalactic space and the foreground Galactic RM.
Observations with very high resolution and at high frequencies
 often help us to understand the intrinsic properties of
radio sources \citep[e.g.][]{ags12,ogg11,tz10,tfs+05}.
On the other hand, observations with a low angular resolution at
low frequencies suffer from differential Faraday rotation {as well
as} internal and external Faraday dispersion
\citep[e.g.][]{sbs+98}, and probe a much shallower {`}skin layer'
of the radio source. This has been found{,} for example{,} in M51
and other nearby galaxies \citep{fbs+11,hbe09,bhb10,bml12}. The
polarized emission from such a shallow layer more often gets the
Faraday rotation in the intergalactic medium between the source
and us.

Therefore, in this paper, we compile the RMs of point-like sources,
unresolved by observations with resolution{ lower} than $1''$ so that
the RMs are mostly produced by{ the} medium between the source of the
emission and the observer, rather than dominated by intri{n}sic
RMs from the sources. We do not collect the RMs of well-resolved
sources 
with a resolution better than $1''$, for which the observed RMs
are mostly intrinsic to the source. If an object has two
components, and each component has a measured RM, we include them
in our catalog as two sources. If a source component is resolved
in polarization observations, we only{ include} the average RM of
the component{ in our catalog} \citep[e.g.][]{prm+89}.

We compiled a catalog of RMs for 4553 sources, as shown in
Table~\ref{tab1}, 
  which are ordered in{ terms of} Right
Ascension (J2000), by searching RMs that have been published in
the literature after{ the} 1980s. Earlier RM compilations by
\citet{skb81} and \citet{bmv88} have been included in our catalog
directly. We may miss a small number of RMs by sporadic
observations for individual objects, and will add new RMs to our
catalog once they become available to us. In the following, we
explain how we compiled our RM catalog{ in more detail}.

 \noindent{\bf Sources with multiple RM measurements:} When two
or more {RMs} are available for one source, we {only }choose one
as the formal RM value of the source by considering{ uncertainties
in} the measurement and the number of observing frequencies,
though in our catalog we list other RM values of the source for
reference. We adopt a formal RM value in the following way.

First of all, we check the source positions. For{ the} current
density of polarized radio sources on the sky, we assume that any
sources within 3$''$ are the same source, because almost all
sources in our catalog were observed with a resolution of $>2''$
except for a few percent (since we set the 1$''$ criterion above).
Many old observations were made with {a }very low resolution of
several arcminutes for very strong sources but no {value for the
}resolution is given in literature, for which we mark them with
``?''  for the resolution in our catalog and check NED (NASA/IPAC
Extragalactic Database){,} NVSS or other catalog for the position,
see below.
Second, for a given source, if the ionospheric RM was carefully
corrected during observation{s}, we take the RM as {a
}formal value.
We then check how many frequency bands or channels were used to derive
the RM. The RM value derived from observations with more frequencies
or channels is more reliable and in general good at removing the $n\pi$
ambiguity. We prefer to take the formal RM from observations with more
channels or bands.
Finally, we look at the uncertainty {in} RMs. The RM with a smaller
uncertainty from{ a} wider{ }frequency range or more sensitive observations
is preferably taken as the formal value.
If two RM values are consistent with
2$\sigma=\sqrt{\sigma_1^2+\sigma_2^2}$, {w}here $\sigma_1$ and
$\sigma_2$ are the uncertainties of the two RM measu{r}ements, we
take the one with {the }small{er} uncertainty. If they differ by
more than $2\sigma$, we take the formal RM derived from the
observations with 
  more channels considering the similar
emission region and Faraday depth for many channels in one band.
If the same RM value and uncertainty for a source appear{s} in two
references, especially when the later authors cite the RM value
obtained by the former authors, we give credit to the former authors.

In Table~\ref{tab1}, we list the formal RM as one entry indicated
by the number of the source, 
 together with other RMs for
this source indicated by ``\_1'', ``\_2'' or ``\_n'' in the
numbering.

 \noindent{\bf Coordinates:} In Table~\ref{tab1} we use the
following labels to indicate how we obtain source positions:

`O': We take the original published positions in the reference of RM
observations;

`M': If coordinates were not published together with the RMs, we
measured source positions from the images or figures in the papers
where the RMs were published;

`N': In some papers, RMs are given for a list of sources with only
object names without coordinates. We checked the
NED \citep[]{hms+91,hms+95} for positions. If observations for RMs
have a high resolution, we assume that the source is associated
with the core, and then take the object position{ that is given in
NED}. If the coordinates of the sources {were} misprinted in a
paper, or if the uncertainties in the published coordinates were
large, we also used the positions from NED.

`V': Some early RM observations have a low angular resolution, and RM
data were listed by{ only} specifying {names of }source{s} without
coordinates.  For these objects we first find the associated NED
objects, then we take the coordinates from the NVSS \citep{ccg+98}.

`F': For some sources RMs have been derived for more than one
component. If the coordinates of each of these components are not
given in the references, we check and take the source positions from
the VLA Faint Images of the Radio Sky survey \citep[FIRST,][]{bwh95}.

 \noindent{\bf Uncertainty level and ionospheric correction:}
We rank the RM uncertainty into {four} levels: A for uncertainties
smaller than 0--3 rad~m$^{-2}$, B and C for 3--10 rad~m$^{-2}$ and
10--30 rad~m$^{-2}$, respectively, and D for uncertainties larger
than 30 rad~m$^{-2}$. In some publications
\citep[e.g.][]{sod+98,rdf+08,mmm+09}, the RM uncertainty was not
given. For those cases we check the figures or position angle
data, and sometimes calculate RM values and their uncertainties
from polarization angle data given in a paper.

The RMs from the ionosphere in general have not been discussed in most
previous papers, except{ for} a few \citep{kkd+94,ow95,ms96,bre08}{, for} which we
should thank {the }careful authors. The combination of the electron density
and the magnetic field in the Earth's ionosphere causes different RMs
for different directions on the sky, especially at sunrise or sunset
\citep{ssh+13}. It may exceed $\pm$5 rad~m$^{-2}$. If the{ contribution of} ionospheric
RM was not mentioned in a paper{,} we interpreted that the
RMs in the paper have not been corrected for the ionospheric RM.  Not
correcting for the ionospheric RM may induce a systematic error for RM
values, which{ in general} should be $2\sim3$ rad~m$^{-2}$. If the
ionospheric RM correction was explicitly made for RM measurements in
the literature, we added ``+'' after the uncertainty level
(i.e. ``B+'').

 \noindent{\bf Object type and redshift:} We checked{ the names
of} object{s} (galaxies or quasars, etc.) for radio sources or
source components. By cross-correlating source positions with
$3''$, 
with NED and
SIMBAD \citep[the Set of
  Identifications, Measurements and Bibliography for Astronomical
  Data:][]{woe+00}, we found object types and redshifts. When NED and
SIMBAD give different object types or redshifts for a given
source, we use the information from NED.

\begin{figure}[h]
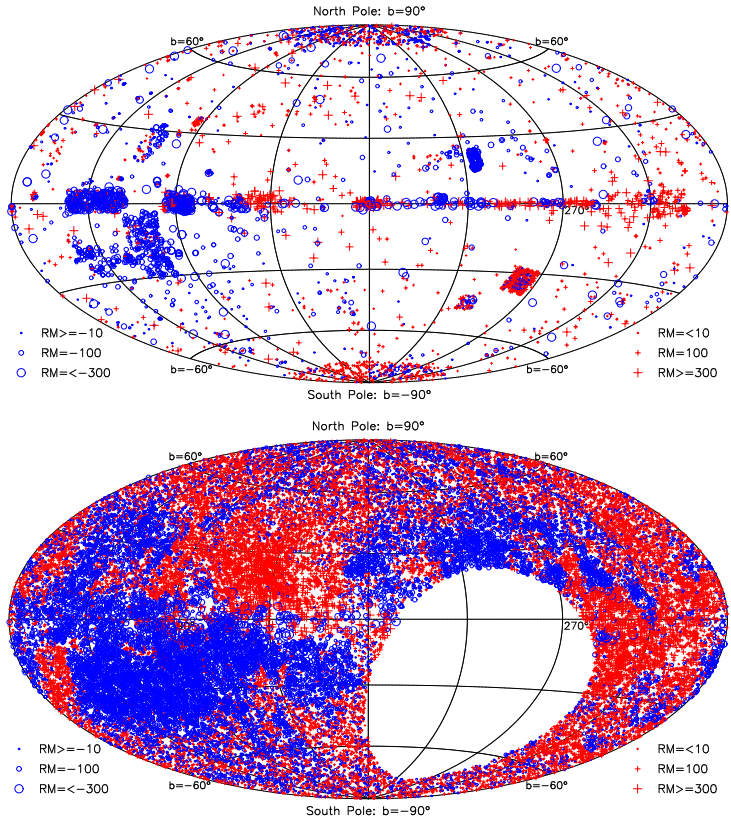


\vs \centering
\includegraphics[angle=-90,width=96mm]{fig1a.ps}

\vs
\includegraphics[angle=-90,width=96mm]{fig1b.ps}

\caption{\baselineskip 3.6mm The sky distribution of the
 compiled RMs in the Galactic coordinates 
   ({\it upper
  }) and that of the NVSS RMs ({\it lower
  }). The linear sizes of the symbols are proportional to the
  square root of the RM values with limits of $\pm10$ and
  $\pm300$~rad~m$^{-2}$. Red pluses indicate pos{i}tive RMs, while blue
  circles indicate negative RMs.  }
\label{RMsky}
\end{figure}

The sky distribution of the compiled RMs is shown in
Figure~\ref{RMsky}. 

\section{Uncertainty {in} the compiled RMs and comparison with the NVSS RM catalog}

\citet{tss09} have reprocessed the NVSS polarization data of the two
IF bands at 1435 MHz and 1365 MHz, and derived RM values for
{37\,543} sources from the two-band polarization data. The sky
distribution of the NVSS RMs is also shown in Figure~\ref{RMsky}.
Our compiled catalog contains RMs of 4553 sources derived from
polarization observations at at least {three} frequencies. Over most
of the sky, the RM distribution is sparse. The NVSS RM catalog
obviously has the advantage of having a large number of sources and
almost uniform sky coverage above a declination of $-40\dg$.

\begin{figure}[b]
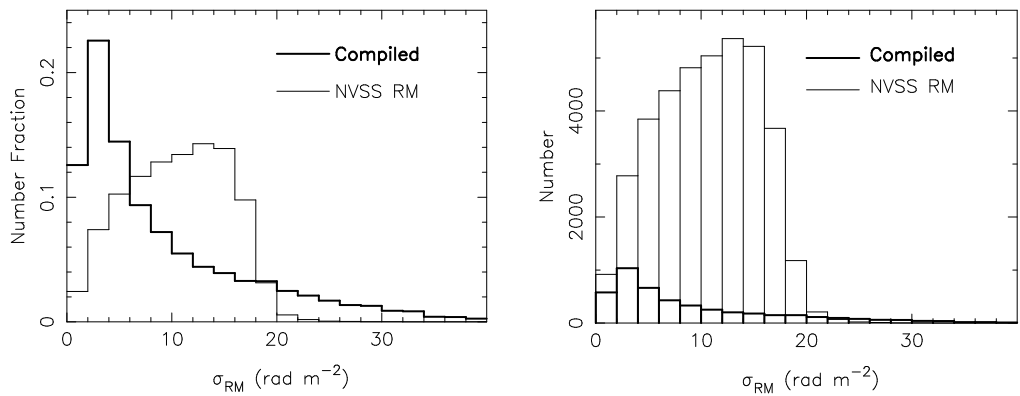


\vs \centering
\includegraphics[angle=-90,width=64mm]{fig2a.ps}~~~~~ \hspace{1mm}
\includegraphics[angle=-90,width=64mm]{fig2b.ps}

\caption{\baselineskip 3.6mm Distribution of the uncertainty of RM
measurements
  $\sigma_{\rm RM}$ for the compiled RM catalog and the NVSS RM
  catalog. The formal uncertainties of the compiled RMs have a peak at
  less than 4~rad~m$^{-2}$, while the uncertainties of the NVSS RMs
  are widely distributed in the range $0-20$\,rad~m$^{-2}$, with a
  peak around 13~rad~m$^{-2}$. Note here that the systematic
  uncertainty of the two RM catalogs ($<3$\,rad~m$^{-2}$ for the
  compiled RMs and 10~rad~m$^{-2}$ for the NVSS RMs) have not been
  considered. }
\label{RMerrsta}
\end{figure}

Here we compare the RM uncertainties in the RM catalog we compiled
with the RM uncertainties from the RM catalog by \citet{tss09}. In
Figure~\ref{RMerrsta} 
 we show the distributions of
the RM uncertainties $\sigma_{\rm RM}$ for the compiled RM catalog
and the NVSS RM catalog. The uncertainties of compiled RMs show a
sharp peak at $\sigma_{\rm
  RM} <4$~rad~m$^{-2}$, because most of the compiled RM data are
determined by polarization angles at more than {three} frequencies
or channels and because the $\Delta \lambda^2$ range is large and
there is no $n\pi$ ambiguity in the data. The uncertainties {in}
the NVSS RMs show a broad distribution with a not-outstanding peak
around $\sigma_{\rm RM}\sim $13~rad~m$^{-2}$, {and} a median
uncertainty{ of} $\sim 10.8$ rad~m$^{-2}$ \citep{sch10,sts11}.

\begin{figure}
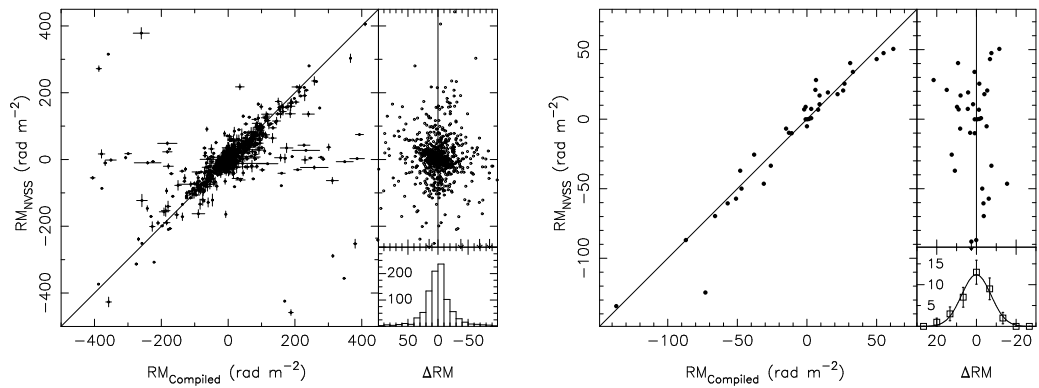


\vs \centering
\includegraphics[angle=-90,width=65mm]{fig3a.ps} \hspace{5mm}
\includegraphics[angle=-90,width=65mm]{fig3b.ps}

\caption{\baselineskip 3.6mm Comparison of RM values for 1024
sources ({\it left}) in both the NVSS
  catalog and the catalog we compiled 
  and for the
  36 sources ({\it right
  }) in both catalogs with formal uncertainties less
  than 1~rad~m$^{-2}$. Notice that the 36 sources
  are all bright sources and that their RMs have been well determined
  in the literature. The $\Delta$RM values of these 36 sources follow
  a Gaussian distribution with a width of $10.46\pm1.45$~rad~m$^{-2}$,
  which mostly comes from the systematic uncertainty of the NVSS RM
  measurements.}
\label{RMcom}
\end{figure}

We also compared the RM values of 1024 sources that appear in both
RM catalogs. In general, most RMs are consistent with each other
within 20 rad~m$^{-2}$ (see Fig.~\ref{RMcom}), though the
distribution of RM difference $\Delta$RM extends to
50~rad~m$^{-2}$, and a few sources even have differences up to
100~rad~m$^{-2}$ \citep[see also fig.~3
  of][]{ptkn11}.

The systematic uncertainties {in} RM data should{ be} but were
rarely recognized in literature. The RMs compiled from the
literature may have not{ been} corrected for the ionospheric RM,
which causes a systematic uncertainty, {at most} 3~rad~m$^{-2}$.
We selected the RMs of 36 sources in both RM catalogs which have
formal uncertainties less than 1~rad~m$^{-2}$, and checked their
RM differences from the NVSS RMs. Because these sources are in
general very bright{ and} the RMs in the compiled RM catalog were
well determined, the distribution of $\Delta$RM must come from the
systematic uncertainty of the NVSS RM catalog. We fit the
distribution with a Gaussian and {obtained} a characteristic width
of $\sigma_0=10.46\pm1.45$ rad~m$^{-2}$ (see the right panel of
Fig.~\ref{RMcom}). Because the upper limit of the systematic
uncertainty from our compiled RMs produced by the uncorrected
ionospheric RM is 3~rad~m$^{-2}$, and even if we discount such{ a}
maximal value{ of} $\sigma_{\rm CM}^{\rm sys}=$ 3~rad~m$^{-2}$,
the systematic RM uncertainty of the NVSS RMs, in add{i}tion to
the formal measurement uncertainty, should be $\sigma_{\rm
NVSS}^{\rm sys}= \sqrt{\sigma_0^2-(\sigma_{\rm CM}^{\rm sys})^2} =
10.0\pm1.5$ rad~m$^{-2}$. This systematic uncertainty can explain
the randomly scattered distribution of a few tens{ of}
rad~m$^{-2}$ around the equivalent line in figure~7 of
\citet{mgh+10}, when the NVSS RMs are compared with the RMs
derived by \citet{mgh+10} for radio sources close to the two
Galactic poles. We therefore agree with \citet{mgh+10} that RMs
derived in \citet{tss09} can be used collectively to describe the
large-scale Galactic RM sky by averaging over large areas, as we
will do in the next section{. However,} one should be cautious to
use the individual NVSS RM values even if they have a very small
formal uncertainty, since these RMs can potentially be inaccurate
due to the systematic{ uncertainty in} RM that we identified. For
example, two standard calibration sources, 3C286 (J133108.3
+303033, with expected RM = 0~rad~m$^{-2}$) and 3C138 (J052109.9
+163822, with expected RM = $-1$~rad~m$^{-2}$), have RM values of
$8.8\pm0.1$ and $7.0\pm0.2$~rad~m$^{-2}$ in the NVSS RM catalog,
respectively. These values are{ only} understandable if such a
systematic uncertainty of the NVSS RM catalog is taken into
account.

\section{The Galactic foreground RM}

We can derive the Galactic foreground RM from all available RM
data. The observed RMs consist of the RM contributions from the
polarized sources themselves, i.e. the intrinsic RM, the RM from
intergalactic space and the RM from the interstellar medium in our
Milky Way. The RM averaging process of a set of sources can smear
out the random RM contributions from intergalactic space, which
are not known exactly but could be random with an amplitude of a
few rad~m$^{-2}$ according to simulations by \citet{ar11} and
recent studies by Xu \& Han (2014). Because the common RM
contribution of a set of neighboring radio sources comes from the
interstellar medium in our Milky Way, the mean RM sky is therefore
{an} excellent description of the Galactic foreground RM. In this
averaging process, we should not use RMs that deviate
significantly from neighboring RMs because those RMs are probably
dominated by the RM contribution that is intrinsic to the source.

\begin{figure}

\vs \centering
\includegraphics[width=104mm]{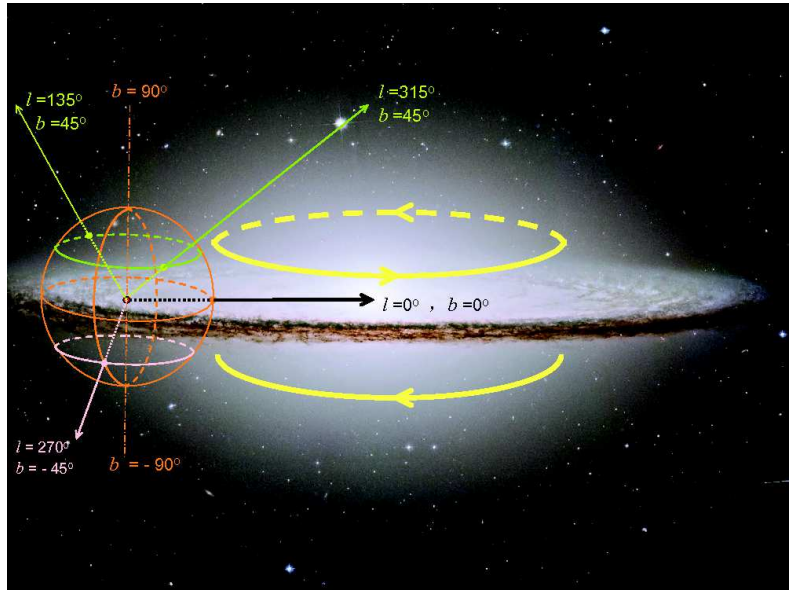}

\caption{\baselineskip 3.6mm The Galactic coordinates and azimuthal
magnetic fields in the
  Galactic halo with reversed directions above and below the Galactic
  plane. Such halo magnetic fields can explain the antisymmetric RM
  sky in the inner Galaxy shown in Fig.~\ref{RMsky} as proposed
  originally by Han et al.\ (1997).}
\label{haloB}
\end{figure}

Many authors have investigated structures in the RM sky, e.g.
\citet{sk80, hmbb97,hmq99,tss09,sts11,ojr+12}. As shown in Figure
\ref{RMsky}, the RM distribution shows large-scale coherent
structures on scales of up to tens of degrees. The most striking
feature is the antisymmetric (quadrupole-like) RM structure in the
inner Galaxy with respect to the Galactic plane and the Galactic
meridian at $l\sim0\dg$, as discussed by \citet{hmbb97,hmq99},
which consists of positive RMs in the upper left and lower right
quadrants and negative RMs in the upper right and lower left
quadrants. Such a pattern has been attributed to large-scale
toroidal magnetic fields in the Galactic halo that go in opposite
directions above and below the Galactic plane (see
Fig.~\ref{haloB}), 
 originally pointed out by
\citet{hmbb97,hmq99} and later modeled by
\citet{ps03,srwe08,ptkn11,jf12b,ft14}. In the two Galactic pole
regions, RMs have much smaller values than {those} near the
Galactic plane, as the magnetic fields in our Milky way are
dominated by the azimuthal components parallel to the Galactic
plane \citep{hmq99,mgh+10}. Small scale structures in the RM sky
are related to known foreground objects, such as HII regions
\citep{hmg11,rgt12}, SNRs \citep{kim88,sim92,srw+11} and high
velocity clouds \citep{mmg+10} in our Milky Way. Here, we do not
study the small-scale RM structure in detail{. I}nstead we derive
the Galactic foreground RM by using the RM catalog we compiled
together with the NVSS RM catalog. We first derive the Galactic
foreground RM, and then compare our result with that obtained by
\citet{ojr+12}.

\subsection{The Galactic Foreground RM and its Uncertainty}

We take all available RM measurements from the compiled RM catalog
and the NVSS RM catalog to estimate the Galactic RM foreground and
its uncertainty. If multiple RMs are listed for a single source,
we only took the formal best value. In total, we have RMs for
{41\,072} sources. Then, any RMs with a formal uncertainty larger
than 30 rad~m$^{-2}$ are discarded, because they are not reliable.
This leaves us {40\,894} sources.

In principle, because of randomness of intergalactic RM
contributions and intrinsic RMs of radio sources, the mean RM of
many radio sources in a small patch of the sky {represents} the
Galactic foreground RM. Because of our position in the Milky Way
as shown in Figure~\ref{haloB} and because of the different
properties of the regular and random magnetic field in the disk
and the halo, the correlation scale of the RM distribution should
be very different in different parts of the sky. The RM
distributions are related {to} each other on large angular scales
for different parts of the halo in the inner Galaxy, {but} the RM
values of background sources vary on small angular scales in the
disk directions because of density fluctuations of the
interstellar medium and reversed fields in the spiral arms.

Previously, there {were} efforts to estimate the Galactic
foreground RM. \citet{fss01} proposed an RM estimation method
which works on a sphere using wavelet approaches; \citet{dc05}
performed spherical harmonic analysis of RM data; \citet{shk07}
put forward Gaussian process convolution models based on the
Markov Chain Monte Carlo method to estimate the Galactic RM
foreground. \citet{ojr+12} proposed to use the signal
reconstruction algorithm for RM sky estimation.  When RM data for
the entire sky are analy{z}ed using spherical harmonics
\citep{dc05} or rely on spatial correlations \citep{ojr+12,jhe04},
RMs from different parts of{ the} sky are assumed to be correlated
in a similar format. We believe that the most secure approach for
deriving the Galactic foreground RM is to calculate the mean RM
from a set of RM measurements of sources in a small patch of sky.

Here we use a simple statistical approach, the weighted average
method, to derive the Galactic foreground RM. Every measurement in
a local sky area around a given line of sight is evaluated and
weighted to calculate the mean, $\langle {\rm RM} \rangle$, and
the uncertainty of the mean, $\sigma_{\langle {\rm RM} \rangle}$:
\begin{equation}
\begin{displaystyle}
 \left\{
\begin{array}{ll}
\langle {\rm RM} \rangle = \displaystyle
\frac{\sum\limits_{i=1}^{N}(w_i {\rm RM}_i)}{\sum\limits_{i=1}^{N} w_i}\,,
\\[3mm]
\sigma_{\langle {\rm RM} \rangle}=\displaystyle \left(
\frac{\sum\limits_{i=1}^{N} w_i({\rm RM}_i-\langle {\rm RM}
\rangle)^2}{(N-1)\sum\limits_{i=1}^{N} w_i} \right) ^{1/2}.
\end{array}
\right.
\label{weighteave}
\end{displaystyle}
\end{equation}
The weight factor $w_i$ is defined as
\begin{equation}
w_i=w_{\sigma_{\rm RM}}\cdot w_{\rm iono} \cdot w_{\rm offset}\, ,
\label{weight}
\end{equation}
where $w_{\sigma_{\rm RM}}$ is the weight for measurement
uncertainty, which consists of not only the formal observational
uncertainty $\sigma_{\rm RM}^{\rm obs}$ but also the systematic
uncertainty $\sigma_{\rm RM}^{\rm sys}$ as we discussed above,
i.e.
$$\sigma_{\rm
  RM} = \Big[(\sigma_{\rm RM}^{\rm obs}) ^2+ (\sigma_{\rm RM}^{\rm
    sys})^2\Big]^{1/2}\, .$$ We adopt the $\sigma_{\rm RM}^{\rm sys} =0 $
rad~m$^{-2}$ for the compiled RMs, and $\sigma_{\rm RM}^{\rm sys}
=10 $ rad~m$^{-2}$ for the NVSS RMs. RM data of sources nearer to
the sightline with better quality play a larger role {in}
determining the Galactic RM foreground at this direction. After
comparisons, we found that in practice $w_{\sigma_{\rm RM}} =
1/\sigma_{\rm RM}^{1/2}$ is superior to $w_{\sigma_{\rm RM}} =
1/\sigma_{\rm RM}$ or $w_{\sigma_{\rm RM}} = 1/\sigma_{\rm RM}^2$
for deriving the Galactic foreground RM, because otherwise RM
measurements with slightly better precision are overemphasized in
the weighted average. The second term is the weight for the
ionospheric RM correction\footnote{Most previous authors
  believe that the ionospheric RM has a small value, within $\pm$5
  rad~m$^{-2}$, and hence{ it is} not worth mention{ing} or correct{ing} in RM
  measurements. However, it is very important to make the ionospheric
  RM correction for many{ fields of} research on{ the} intergalactic medium, and it
  {will }become even more important {during the} SKA era{ in the future}.
  For example, when one
  looks for the residual RM evolution of a few rad~m$^{-2}$ with
  redshift from intergalactic magnetic fields or fields in{ the} cosmic web (e.g. Xu \& Han 2014),
  only RM observations with proper ionospheric RM corrections can
  really reduce such a systematic uncertainty in the estimation of the
  Galactic foreground RMs and ultimately the residual RMs. Therefore,
  RMs with proper ionospheric RM corrections are emphasized here,
  and are given a double weight in Sect.~4.1 for the averaging method
  so that it plays a ``calibration" role. However, because of{ the}
  relatively small number of RMs with ionospheric RM correction, no
  obvious difference can be seen in the final foreground RM map if one
  take{s} $w_{\rm iono}=1.0$ for all RM data.}. If the RM of a source has
been corrected for the ionospheric RM, then we set $w_{\rm iono}=1.0$,
otherwise $w_{\rm iono}=0.5$. The final term is the weight factor
$w_{\rm offset}$ for the angular distance $a$ of a source to the given
line of sight, which is defined as {the} Gaussian function $w_{\rm
  offset}= \exp(- a^2/2a_0^2)$, where $a_0$ is the characteristic
width. RM values of farther sources with a larger $a$ have less weight
in calculations of the Galactic foreground RM for a given
direction. For example, $a=0$, $w_{\rm offset}=1$; $a=a_0$, $w_{\rm
  offset}=0.607$; and $a=2a_0$, $w_{\rm offset}=0.135$. \citet{ow95}
calculated the mean RM for all sightlines within 20\dg, 
assigning equal weights to each of these measurements. Similarly,
\citet{hmbb97} used a radius of $15^\circ$. With the larger
surface density of sightlines in our sample{,} we can choose a
much smaller $a_0=3\dg$. We calculate the mean RM using RM data
within $2a_0$ for any given direction. Over most of the NVSS
region there are at least 10 
measurements within this region. In the southern sky of Dec$
<-40\degr$, however, the RMs are scarce, and we have to increase
$a_0$ from $3\dg$ to $6\dg$ or $9\dg$ or $12\dg$ so that we always
use at least 10 RMs to calculate the average RM. In the future,
when RM data of more radio sources will become available, one can
choose a smaller $a_0$.
We make it possible to change the weighting scheme used to
calculate the Galactic foreground RM on our website.

\begin{figure}[pht]
\vs

\centering
\includegraphics[angle=-90,width=105mm]{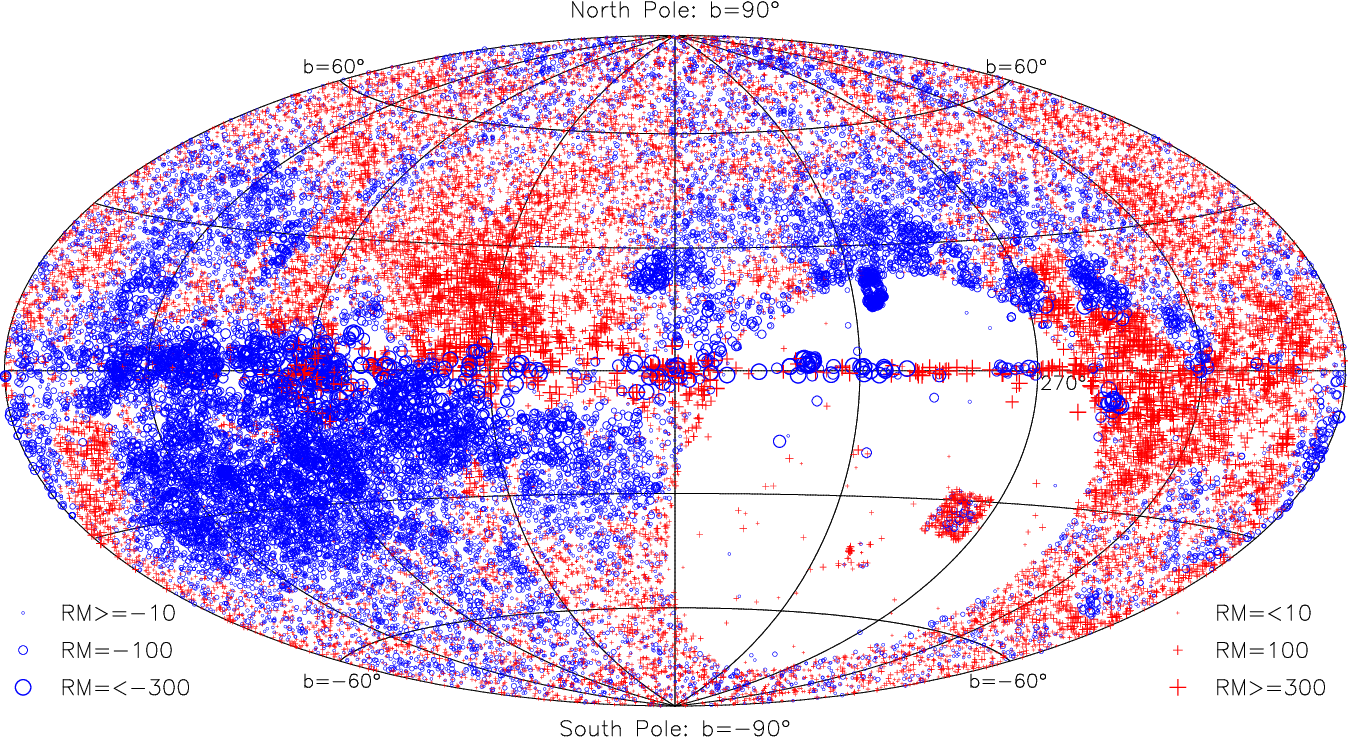}

\caption{\baselineskip 3.6mm The RM distribution for sources from
the compiled RM catalog and the NVSS RM catalog. Outliers and RMs
with an uncertainty larger than 30~rad~m$^{-2}$ have been
discarded.} \label{RMdata}

\vs \centering
\includegraphics[angle=-90,width=95mm]{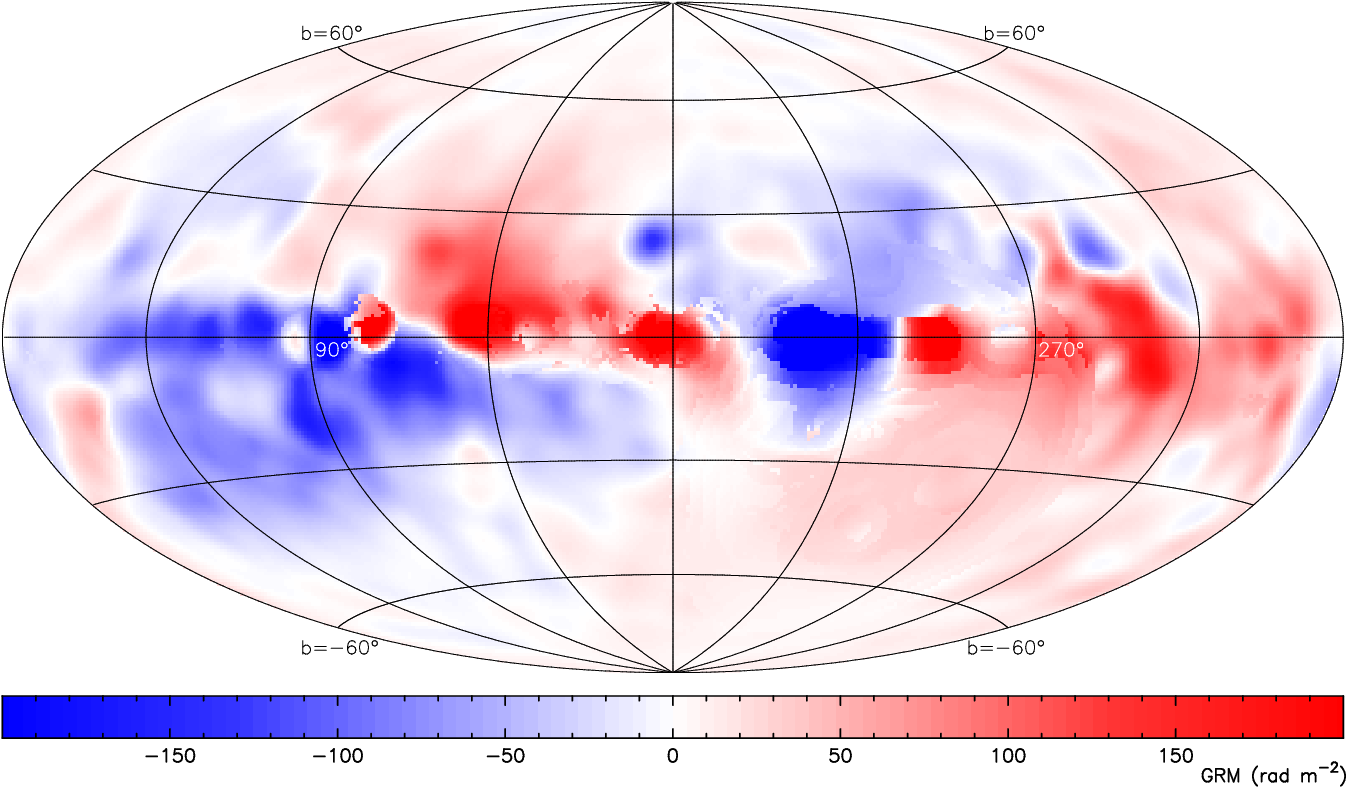}
\includegraphics[angle=-90,width=95mm]{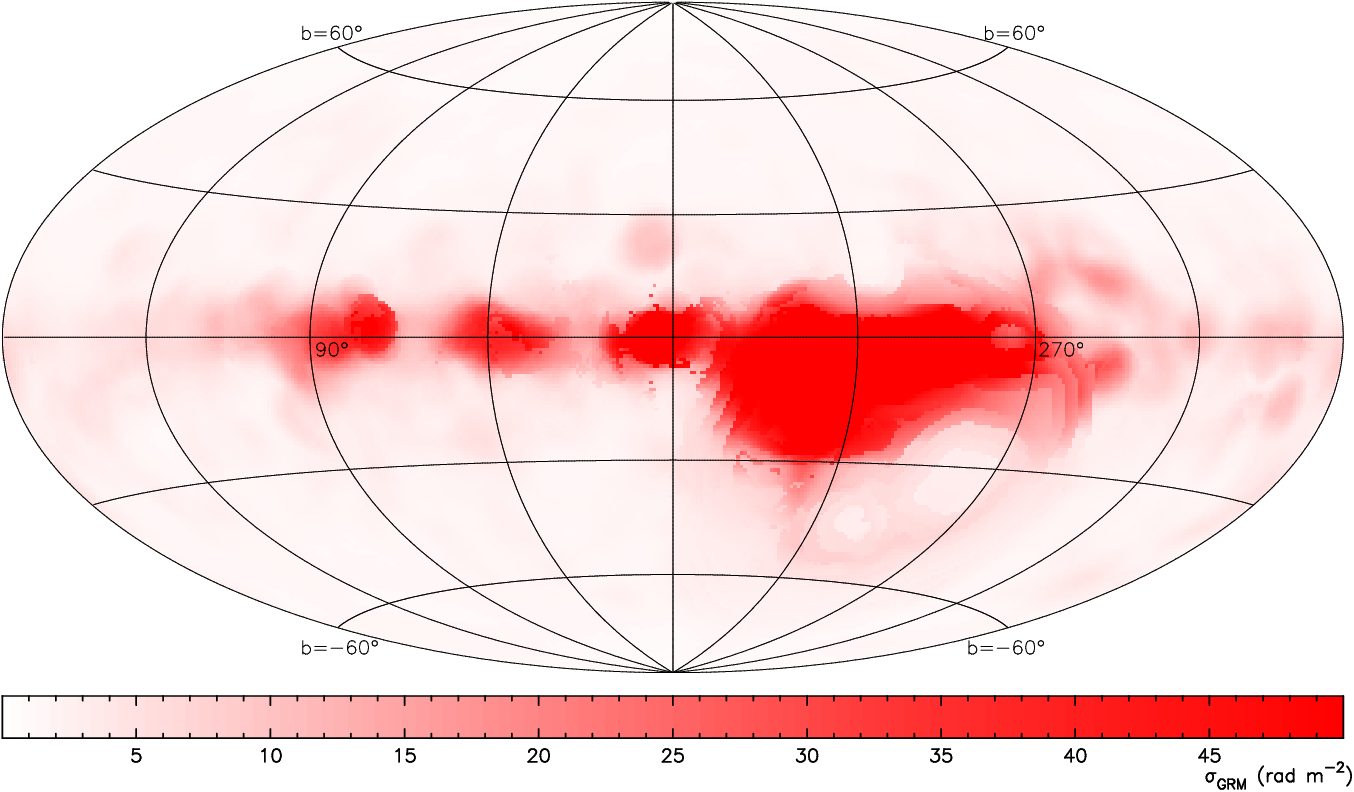}

\vspace{-3mm} \caption{\baselineskip 3.6mm The Galactic foreground
RM map ({\it top panel}) and its associated
  uncertainty map ({\it bottom panel}) that we calculated by combining the
  compiled RM catalog with the RM catalog by \citet{tss09}.
} \label{RMskyme}
\end{figure}

To derive the Galactic foreground RM map, as the first step it is
necessary to filter out the ``anomalous'' RMs if they are
obviously deviating from their neighbors, because such RMs are
almost certainly dominated by intrinsic RMs. Such a filtering
procedure {for} outliers was not done by \citet{ojr+12}, but has
{already }been included in some early work \citep{hmbb97,hmq99}
and recent modeling \citep[e.g.][]{jf12b}. We compare the RM value
of each source with the weighted mean in
{Equation~(}\ref{weighteave}{)} and the weighted standard
deviation
 $$\sigma= \left[\sum_{i=1}^{N} w_i\Big({\rm RM}_i-\langle {\rm
RM} \rangle\Big)^2 \Big/
  \left(\sum_{i=1}^{N} w_i\right)\right]^{1/2}$$
  of neighboring sources within $3\degr${,}
$6\dg${,} $9\dg$ or $12\dg$ as mentioned above, except for the
target source. If the RM of the target source deviates more than
3$\sigma$ from the mean of surrounding sources, then we discard it
as an outlier. Galaxy clusters can contribute large RMs to
background sources \citep{ckb01,bfm+10}, RMs of some radio sources
behind galaxy clusters are ``anomalous{,}'' and hence can be
removed by this step in our analysis. After iterating a few times,
we get good RMs for {39\,984} sources that we can use in our
reconstruction of the Galactic foreground RM (see
Fig.~\ref{RMdata}). 
 The scarcity of RM data is
obvious in the region Dec $<-40\dg$ which is not covered by the
NVSS.

Using these RM data with the outliers removed, and by applying the
weighted average in {Equation}~(\ref{weighteave}), we obtain the
RM map of the Galactic foreground and its uncertainty, which we
show in Figure~\ref{RMskyme}. 
 Small-scale
structures appear near the Galactic plane and towards some HII
regions \citep[e.g. Sh 2-27 at ($l=8.0$, $b=  +23.5$),][]{hmg11},
and the large-scale foreground RM is also visible away from the
Galactic plane. The uncertainty is obviously large in the southern
sky at Dec $<-40\dg$ due to the shortage of RM data. A larger
scatter in the RM data and hence a larger uncertainty in the
estimated Galactic foreground RM is found near the Galactic plane,
especially in the inner Galaxy near tangential directions of
spiral arms, where more turbulent clouds along the line of sight
are expected.

\begin{figure}
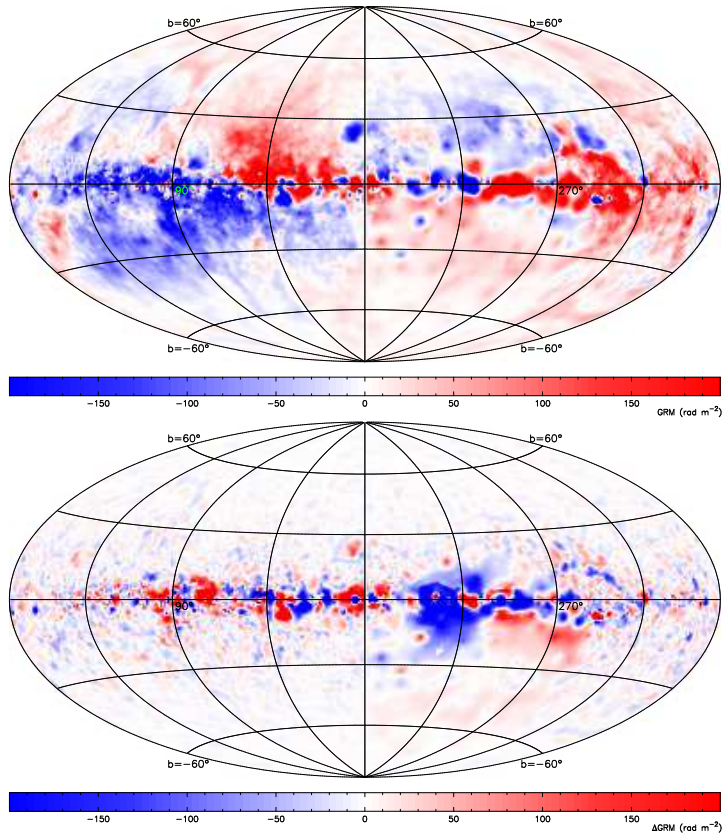


\vs \centering
\includegraphics[angle=-90,width=95mm]{fig7a.ps}
\includegraphics[angle=-90,width=95mm]{fig7b.ps}

\caption{\baselineskip 3.6mm The Galactic foreground RM derived by
\citet{ojr+12} using
  information field theory, and the difference map from our RM
  foreground.}
\label{mapOpp}
\end{figure}

\subsection{Comparison of Our Foreground RM Map with that of \citet{ojr+12}}

At present we have calculated the weighted average RM of the
cleaned RM data for the Galactic foreground RM. Given enough data
points in a small area for averaging, our approach is very simple
and very straightforward, in comparison with previous efforts
\citep{fss01,dc05,shk07,ojr+12}. The latest such an attempt before
our work was made by \citet{ore11,ojr+12} who used {a }signal
reconstruction algorithm based on information field theory
\citep{efk09}. They took into account the spatial correlations and
used the{ formalism of an} {\it extended critical filter}
\citep{ore11} to reconstruct the map for the Galactic foreground
RM (see Fig.~\ref{mapOpp}). 
 Using priors for the signal
$s$, noise $n$, the angular power spectrum and the noise
correction factors, they calculate the mean $m=\langle s \rangle
_{P(s|d)}$ {from data $d$}, i.e. the reconstructed signal by
iterating filtering equations \citep[eq.~(9) -- (11) in
][]{ojr+12}. The posterior mean for the Galactic Faraday depth is
given by $\langle \phi \rangle = pm$. The critical step in the
filtering process is to identify the posterior probability density
based on prior information. The relation between the posterior
mean and measured data contains an information propagator $D$
\citep{efk09} which describes how the information contained in the
data at one position propagates to another position. The filtering
equations are designed in the framework of 
a series expansion in spherical harmonics, where the minimum
length scale $l_{\max}$ is limited by the pixel size of the
discretization. In their final Faraday depth map (see the upper
panel of Fig.~\ref{mapOpp}), there are many small structures. RMs
in some areas that differ greatly from their surroundings come
from outliers or RM values with very large uncertainties (see the
difference map in the lower panel of Fig.~\ref{mapOpp}). The
angular resolution of their ``extended critical filter'' algorithm
seems to be high enough to partially pick up anomalous RM values,
though such a resolution seems to be necessary for recovering
small-scale structures near the Galactic plane if there are enough
RM data. On the other hand, their approach seems to be very good
at extrapolating the foreground RM in the undersampled halo region
at Dec $<-40\degr$ using spherical harmonic components.  In our
approach we only consider RM data close to any given line of
sight, without RM outliers, for the Galactic foreground RM. The
RMs with different uncertainties are simply weighted for the
averaging calculations. The uncertainty in the map of the Galactic
foreground RM that we constructed depends on the number and
quality of RM data points in a local area.

\begin{figure}
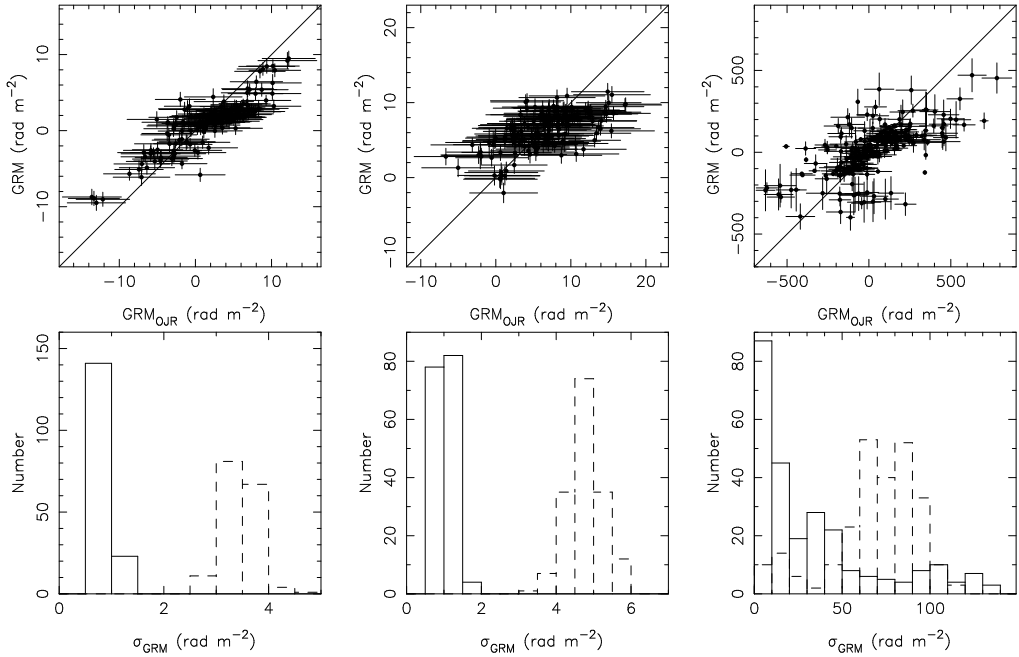


 \centering
\begin{tabular}{ccc}
\mbox{\includegraphics[angle=-90,width=42mm]{fig8a.ps}}&
\mbox{\includegraphics[angle=-90,width=42mm]{fig8b.ps}}&
\mbox{\includegraphics[angle=-90,width=42mm]{fig8c.ps}}
\end{tabular}

\vspace{-2mm} \caption{\baselineskip 3.7mm Comparison of the
Galactic foreground RM calculated by our
  weighted mean and by \citet{ojr+12} in the upper panels for three
  regions near the North Gala{c}tic pole ($b > 75 \degr$, {\it left
    panels}), the South Galactic pole ($b < -75 \degr$, {\it middle panels})
  and the Galactic plane of all Galactic longitude{s} ($b = 0\degr$, {\it
    right panels}). We calculate the Galactic foreground RM for 160
  of 3279 pixels in each polar cap with a separation of pixels larger
  than $1\degr$, and 256 equally separated pixels along the Galactic
  plane. The distributions of uncertainty of the Galactic foreground RM
  are compared in the lower panels, with {a }solid line for our
  calculations and {a} dashed line for the results by \citet{ojr+12}.}
\label{GRMcomp}
\end{figure}

We compare the values of the Galactic foreground RM calculated by
our method and by \citet{ojr+12} towards the North Galactic Pole,
{the }South Galactic Pole and the Galactic plane (see
Fig.~\ref{GRMcomp}), 
 and {find} that they are more or
less consistent. However,{ in general} our approach gives more
reliable estimates of the Galactic RM foreground with smaller
uncertainties by using more than 30 sources for averaging. The
uncertainty for pixels near the Galactic plane is dominated by
scatter in the RM data.

\section{Concluding remarks}

We compiled a{n} RM catalog of 4553 sources which have a small
systematic uncertainty. Even though the NVSS RM catalog by
\citet{tss09} contains {37\,543} RMs, the measurement
uncertainties of their RMs are large and the RM values suffer from
an additional systematic uncertainty of $10.0\pm1.5$ rad~m$^{-2}$.
The RM catalog we compiled provides a {database}
for future calibration or comparisons with wideband
observations.

We make all compiled RM data publicly available on this webpage:
{\it http://zmtt.bao.ac.cn/RM/}, and provide an interface on this
webpage to extract the RM data for a region and to calculate the
Galactic RM foreground. The RM data can be downloaded from the
webpage. We will continuously update the RM catalog on our webpage
by including newly published RM values. Knowing the RM of the
Galactic foreground is important for many research fields, such as
magnetic fields in galaxy clusters \citep[e.g.][]{bvb+13},
Galactic bubbles \citep[e.g.][]{ssf13}, HII regions
\citep[e.g.][]{hmg11} and {SNRs} \citep[e.g.][]{srw+11}. Using the
RM catalog that we compiled together with the NVSS RM catalog,
users can always get the best estimates of the Galactic foreground
RM for any direction {i}n the sky by using a weighted averaging
method.

Finally, we would like to remind the users of the compiled RM
catalog to also cite the original RM observation papers
 if any individual  RM data are used.

\normalem

\begin{acknowledgements}
We thank the referee for very careful reading and very detailed
comments, Dr.  Hui Shi for kind helps in the RM collection from
some papers, Dr. Tracy Clark for providing RM data {from} her PhD
thesis and Dr. Bin Cui for his helps and advice in development of
the webpage for the RM catalog. J. Xu would especially like to
thank X. Y. Gao and P. F. Wang for helpful discussions.
The authors are financially supported by NAOC135 Grants and the
Strategic Priority Research Program ``The Emergence of
Cosmological Structures'' (Grant No. XDB09010200) of the Chinese
Academy of Sciences.
This research has extensively used the {databases} of the National
Radio Astronomy Observatory VLA Sky Survey (NVSS), VLA Faint
Images of the Radio Sky (FIRST), the set of Identifications,
Measurements and Bibliography for Astronomical Data (SIMBAD) and
the NASA/IPAC Extragalactic {Database} (NED) which is operated by
the {Jet Propulsion Laboratory}, California Institute of
Technology, under contract with the National Aeronautics and Space
Administration.
\end{acknowledgements}

\end{document}